\documentclass[journal]{IEEEtran}
\usepackage{graphicx}        
\usepackage{amsmath}    
\usepackage{amsthm}                         
\usepackage{multirow}
\usepackage{enumerate}
\usepackage{latexsym}
\usepackage{amssymb}
\usepackage{txfonts}
\usepackage{caption}
\begin{document}
%
\title{Cooperative Spectrum Sensing Scheduling in Multi-channel Cognitive Radio Networks: A broad perspective}
%
%
%

\author{Prakash~Chauhan, Sanjib~K.~Deka, Monisha Devi,
        and~Nityananda~Sarma~\IEEEmembership{Member,~IEEE}        
\thanks{P. Chauhan, S. K. Deka, M.Devi and N. Sarma are with the Department
of Computer Science and Engineering, Tezpur University, Tezpur, Assam, 784028, India 
e-mail: \{prakashc, sdeka, magna, nitya\}@tezu.ernet.in} 
}

\maketitle

\begin{abstract}
Cooperative spectrum sensing has been proven to improve sensing performance of cognitive users in presence of spectral diversity. For multi-channel CRN (MC-CRN), designing a cooperative spectrum sensing scheme becomes quite challenging as it needs an optimal sensing scheduling scheme, which schedules cognitive users to different channels such that a good balance between detection performance and the discovery of spectrum holes is achieved. The main issue associated with cooperative spectrum sensing scheduling (CSSS) scheme is the design of an optimal schedule that could specify which SUs should be assigned to which channels at what time to achieve maximal network throughput with minimal amount of energy and satisfying the desirable sensing accuracy. In this regard, designing an efficient CSSS scheme for MC-CRN is of utmost importance for practical implementation of CRN. In this article, we explore CSSS problem from a broad perspective and present the different objectives and the inherent issues and challenges arise in designing the CSSS scheme. We also discuss the different methods used for modeling of the CSSS scheme for MC-CRN. Further, a few future research directions are listed which need investigation while developing the solution for CSSS problem.

\end{abstract}

%

\section{Introduction}
%
%
%
%
\IEEEPARstart{T} he seamless and reliable wireless communication has been the high demand in present time. Recently research in cognitive radio (CR) based networks has drawn attention from universities, organizations due to its ability to solve the potential problem of spectrum scarcity \cite{1Akylidiz-06}. At the forefront the IEEE standard 802.22 called Wireless Radio Area Network (WRAN) \cite{3IEEE}, DARPA's NeXt Generation project \cite{2DARPA}, Microsoft and Google's White Space Coalition (WSC) are some of the major projects in CR Network (CRN) which envision the future opportunistic wireless technology. CR is the key technology for dynamic spectrum access (DSA) with capability to use licensed wireless channels in an opportunistic manner by means of discovering the spectrum holes. A spectrum hole is the portion of licensed spectrum band which remains unused by its licensed users. A CRN is a network consisting of CR enabled nodes (secondary users) and the licensed nodes (primary users), where the CR enabled nodes share communication in licensed spectrum bands by utilizing spectrum holes. The detection of spectrum holes in licensed bands by the cognitive users or secondary users (SUs) is important for the success of CRN and this functionality is termed as spectrum sensing. The spectrum sensing functionality is typically realized using two statistical hypotheses $H_1$ and $H_0$ which indicate the presence or absence of primary user signal in the licensed band. Usually spectrum sensing techniques (or algorithms) are associated with two probabilities value: probability of detection $(P_d=P[H_1|H_1])$ which indicates the probability of correctly detecting the presence of primary signal when the signal is actually present and probability of false alarm $(P_f=P[H_1|H_0])$ which indicates the probability of falsely declaring the presence of primary signal when the signal is actually absent. 
 Higher values of $P_d$ and lower values of $P_f$ are always desirable for SUs, because higher $P_d$ value ensures the minimum possibility of interference to PU transmission and lower $P_f$ value makes better possibility of throughput to be attained by SUs. Therefore, the sensing performance of a SU is constrained by the values of $P_d$ and $P_f$ which are again dependent on the characteristics of the licensed spectrum band. The characteristic of licensed spectrum at different point in time and varied geographical locations is referred to as spatio-temporal characteristics. 

 \begin{figure}[!]
 	\begin{center}
 		\includegraphics [width=8.5cm, height=6cm]{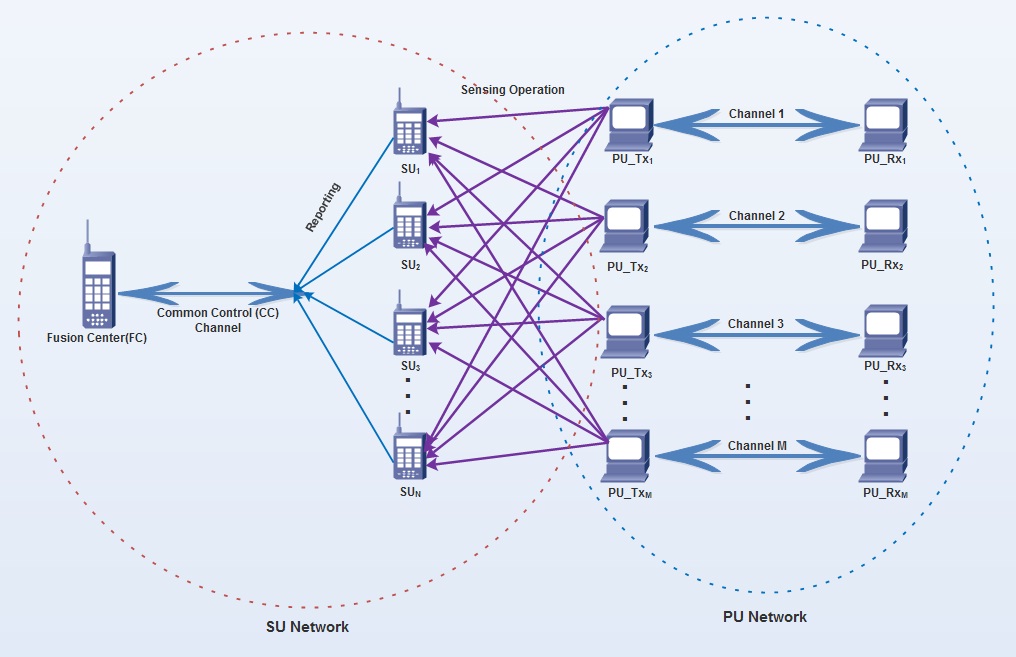}
 		\caption{Framework of Cooperative Spectrum Sensing}
 	\end{center}
 	\label{F:Figure1}
 \end{figure}
\par In CRN, SUs are equipped with CR capabilities like frequency agility, adaptive modulation, transmits power control and can cope up dynamically to perform spectrum sensing. With the ability to sense spatio-temporal variations in the radio environment a SU can find and exploit unoccupied spectrum portions through dynamically adjusting operating frequency, bandwidth and physical layer parameters. In this regard the accuracy of single user sensing performance suffers due to the issues arising from spectral diversity like multipath fading, shadowing, hidden terminal problem and receiver uncertainty problem \cite{4AkylidizBala-11}; which can be mitigated by cooperative spectrum sensing (CSS) technique (or scheme) in which spatially diverse SUs cooperate and collaboratively make a sensing decision. As depicted in Fig. 1 the basic framework of a CSS scheme is consists of groups of SUs to perform sensing of PU channels and a fusion center to take a decision in collaborative manner where SUs communicate with fusion center using a common control channel. While carrying out CSS, the SUs in a group synchronize among themselves using a slotted time system as shown in Fig. 2 where each of the slot is further subdivided for sensing, reporting and transmission purposes. 
\begin{figure}[!]
	\begin{center}
		\includegraphics [width=8.5cm, height=4.5cm]{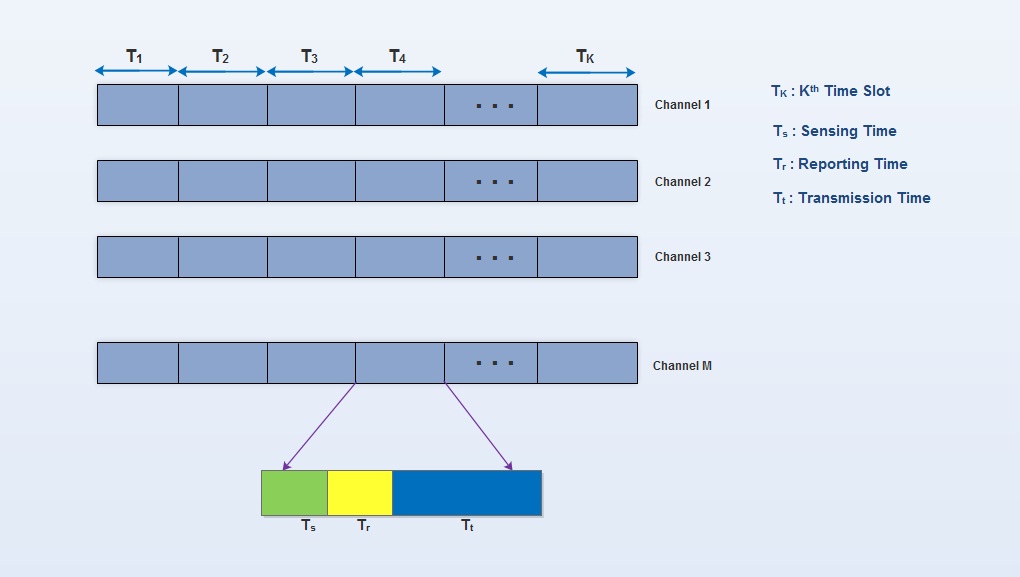}
		\caption{Structure of time slot}
	\end{center}
	\label{F:Figure2}
\end{figure}
\par 
Due to varying spatio-temporal characteristics of licensed channels, there may exist a large number of spectrum holes. Exploiting such spectral opportunities or holes in multiple licensed channels improves spectral efficiency. Taking the challenge to detect such spectrum holes in MC-CRN through CSS on multiple channels, needs an optimal sensing scheduling scheme which can appropriately assign SUs to different channels such that a good balance between detection performance and the exploration of spectrum holes is achieved. The main issue associated with CSS scheduling (CSSS) scheme is the design of an optimal schedule that could specify which SUs should be assigned to which channels at what time to achieve maximal network throughput with minimal amount of energy and satisfying the desirable sensing accuracy. Furthermore the scheduling problem faced in cooperative sensing for MC-CRN is a combinatorial optimization problem and is shown to be NP-hard \cite{nphard2010song}. Therefore, designing an efficient CSS scheduling scheme for MC-CRN is of utmost important for practical implementation of CRN. In this article, we explore CSSS problem from a broad perspective and present the different objectives and the inherent issues and challenges arise in designing the CSSS scheme. We also discuss the different methods used for modeling of the CSSS scheme for MC-CRN. Further, a few future directions are listed which need investigation while developing the solution for CSSS problem. From the best of our knowledge this article is the first one to report the different aspects of CSSS problem for MC-CRN environment.  


\par The rest of the article are organized as follows. Section II presents a basic framework of CSSS. In Section III the objectives of CSSS schemes are presented while issues and challenges faced during CSSS modeling are being discussed in section IV.  Methods for modelling CSSS scheme are highlighted in section V and future directions are discussed in Section VI. Finally we draw the conclusions in Section VII.
\section{CSSS Framework}
	\begin{figure}[!]
		\begin{center}
			\includegraphics [width=8.8cm, height=4.5cm]{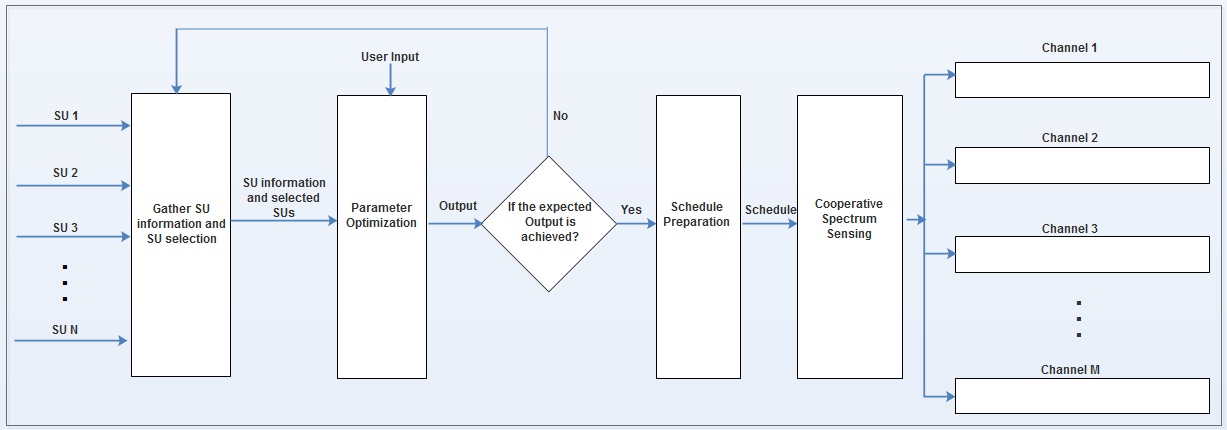}
			\caption{ A basic framework for Cooperative Spectrum Sensing Scheduling Scheme}
		\end{center}
		\label{F:Figure3}
	\end{figure}
	
A basic framework for CSS scheduling scheme is shown in Fig. 3. The building blocks of the framework depict the set of operations performed during CSS scheduling. At the initial block information about all the SUs are gathered and a subset of SUs is selected. The SU information includes individual $P_d$, $P_f$, received SNR of PU channels, and energy constraint. The selected SUs along with their information is passed to the next block where parameter optimization is performed. In this block the user inputs which include targeted $P_d$ and $P_f$ for the particular channel, energy constraint and channel characteristics information are taken into account during optimization of the parameters. The parameter optimization operation decides an optimal set of SUs for each channel by tuning parameters available to this block. After the optimization of parameter if desired output is achieved then the output information is passed to the next block otherwise a new set of SUs are selected and the process is repeated. The schedule preparation block prepares a schedule which contains the information regarding which SUs will sense which channels, in what order, for how long, and at what time. Once the schedule is prepared the next block i.e. cooperative spectrum sensing block starts sensing operation by deploying the selected SUs on the designated PU channels. 

\section{CSSS Objectives}
Broadly the main objectives of developing CSS scheduling scheme for MC-CRN are elaborated as follows:
\subsubsection{Improvement of detection performances as well as throughput}
The goal of CSSS is to improve the detection performance in terms of better $P_d$ and $P_f$ while sensing multiple licensed channels. Improving the $P_d$ enhances PU protection by means of resulting least interference to PU communication and hence penalty gets minimized. Reducing the $P_f$ increases the chances of using free channels for SU transmission which eventually increases the SU throughput. Again spending less time in channel sensing increases duration of transmission which ultimately leads to better throughput. 
\subsection{Energy efficiency}
In MC-CRN while designing scheduling scheme for CSS, the sole consideration of detection performance is not always feasible due to limited battery life of SUs. In such scenario maintaining a trade-off between detection performance and energy efficiency becomes very important for SUs to prolong the network lifetime. An efficient scheduling strategy can effectively decide the order of sensing in which SUs perform CSS as well as the list of channels in which sensing should be done so that the overall performance of the network get increase and overall energy consumption by the SUs get minimize.
\subsection{QoS provisioning} 
Quality of Service (QoS) is always a matter of concern in wireless communication because of vulnerability of wireless communication to different environmental hazards. 
During CSS scheduling in MC-CRN QoS plays important role from both the primary and secondary users’ perspective. The objective of PU QoS indicates the requirement of sensing accuracy of SUs such that the SUs cannot disrupt the PU communication. The objective of SU QoS indicates that while a SU indulge in spectrum sensing, the channel that is sensed to be free should be usable for SU transmission. Again it is important that the expected idle duration of the channel detected while sensing should be able to satisfy the overall spectrum need of all SUs.
\subsection{Fairness consideration}
The objective of fairness consideration becomes crucial while dealing with heterogeneous SUs in MC-CRN. This is because in heterogeneous environment different SUs might have different energy level and they might be exposed to different set of channels with varying channel characteristics. In such scenario, while sensing the licensed channels, some SUs might waste more energy than the other due to the poor link SNR. This is because low link SNR force SUs to sense longer duration which eventually increases energy consumption. As a result of this some of the SUs might get their battery drained and may not be able to communicate during transmission slot. So while designing an efficient CSSS scheme, the consideration of fairness amongst the SUs is very important so that every SU contribute fairly during sensing and get a fair utility during communication.


\section{Issues and challenges}
In designing CSSS schemes for MC-CRN there are several issues and challenges which need to be mitigated during the problem formulation. The major issues and challenges in this context can broadly be classified as follows: 
\subsection{Trade-off between sensing accuracy and spectral opportunities:}
The success in opportunistic channel usage by CRN depends on how the SUs are deployed to explore the spectrum holes available in multiple licensed channels such that the trade-off of spectrum sensing accuracy requirement and spectral opportunities available always remain balanced. In MC-CRN the sensing accuracy expected is always high along with the spectrum opportunities being fully exploited. High sensing accuracy is important to protect the PU transmission from harmful interference. On the other hand to ensure better spectral utilization, SUs need to explore more and more numbers of channels. The challenge in MC-CRN is how and in which order to explore multiple licensed channels by SUs, provided PU detection and false alarm constraints that are maintained during the sensing functionality. Again maintaining the balance of the trade-off of sensing accuracy and spectral opportunities available depends on number of SUs and their scheduling to perform the cooperative spectrum sensing functionality. If more and more SUs are assigned to perform CSS for a single or fewer numbers of licensed channels the sensing duration required will be substantially reduced as shown in Fig. 4 which eventually increases the throughput as the SUs will get more transmission time. Further the sensing accuracy will be improved as more numbers of SUs will cooperate for PU detection decision. On the contrary the possible spectrum opportunities available in all other licensed channels will not be fully exploited since fewer numbers of licensed channels are sensed by the SUs. Therefore to fully exploit the spectral opportunities selected subset of the SUs need to be assigned to perform the CSS for all the available licensed channels. In this regard the CSS technique needs to decide about which SUs are to be selected to perform CSS and for which particular licensed channel. Deciding an appropriate sensing schedule in this context is of utmost importance in multi-channel CSS framework.
\subsection{Heterogeneity of PUs and SUs}
Heterogeneity of PUs and SUs which physically operate in diversified manner is a realistic scenario in CRN. There are several heterogeneity issues which need to be taken into account during the modeling of CSSS framework for MC-CRN. The heterogeneity issues due to possible changes in network environment for primary as well as secondary network counterpart such as, deployment of new PU and SU nodes, the omission or removal of existing PU and SU nodes, and the change in wireless channel condition are important to be incorporated into the algorithm such that SUs can adapt to the realistic environment as much as possible. The licensed channel characteristic in this context in terms of gain, idle probability statistics and channel capacity are different for different channels. Further the PU activity status and channel fading statistics for SUs may not be identical for different channels which needs to be taken into account during the modeling of CSSS scheme \cite{6sun2013energy}. Due to the difference in capability of SUs in terms of sensing performance and energy efficiency because of different factors like hardware issue, how accurately the cooperative sensing framework and sensing schedule can be designed is important for MC-CRN.
\subsection{Imperfect sensing:}
The concept of perfect spectrum sensing assumed in many of the existing works in CRN rely mostly on homogeneous network environment with stable noise condition. But in reality the practical challenge of CRN is to deal with the network environment which may be imperfect due to the fact that in practice there are many primary radio networks (PRNs) available, usually those are heterogeneous by nature and channels are exposed to different fading environment. The modeling of CSS framework and designing the scheduling scheme in such an multi-PRN multi-channel environment require sophisticated approach to take into account the difference in network characteristics such as PU traffic, the number of shared channels, interference tolerance of PRNs and available spectrum holes in multi-PRN channels \cite{imperfectkim2013optimal}. Devising the technique to measure and differentiate such network characteristics which are distributed unevenly in multi-PRN environment is challenging and may lead to imperfect sensing results. Further, in presence of adverse fading and shadowing effects due to different overlapped PRNs, technique to resolve the widely varying degree of sensing capabilities of SUs is required. Further, in presence of such a multi-PRN environment, technique to accommodate the issues of diversity of multi-channel scenario is another important requirement to determine the CSS schedule.  
\begin{figure}[!ht]
	\begin{center}
		\includegraphics [width=8.6cm, height=6cm]{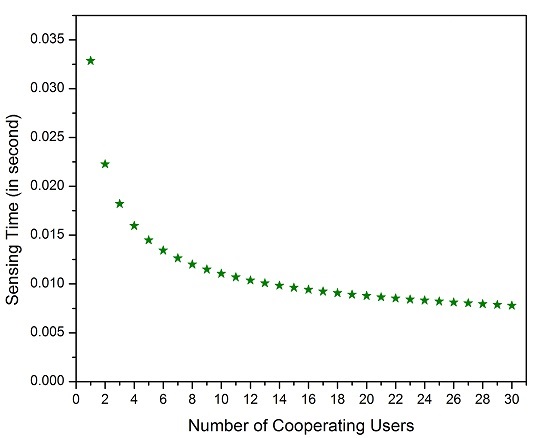}
		\caption{Number of Cooperating Users vs. Sensing Time}
	\end{center}
	\label{F:Figure4}
\end{figure}

\subsection{Implementation Complexity:}

In MC-CRN, a fast opportunity discovery is utmost important for seamless service. Otherwise on wasting more time in opportunity discovery, SUs may limit itself to utilizing the available bandwidth and may not achieve data transmission service with full strength. In this regard to satisfy the bandwidth requirement of application, a SU needs to discover more and more opportunities in different licensed channels. In order to sense such a set of channels deriving an optimal sensing sequence is important such that the latency for discovering the additional opportunities (i.e. the idle channels) can be minimized as much as possible. 
The performance of such algorithms are expected to be comparable at per the optimal results in minimum time and in less computationally complex manner. 
In MC-CRN the complexity of sensing wide (larger than 1 GHz) \cite{complexitywang2014cooperative} range spectrum band arises due to the fact that the spectrum resources owned by different PUs are dynamically wider in nature. Further, the sampling rate requirement of a SU in terms of twice the available bandwidth makes the SUs operating in such wider spectrum bands infeasible due to the issues of computational complexity and hardware limitations. Hindering with limitation of sensing capability to sense wide range spectrum bands and the complexity of network environment deciding the scheme about how to coordinate among the SUs is required to be addressed. 

\subsection{QoS and Fairness:}
Focusing on the application requirements in CRN and to protect QoS parameters of PU as well as SU transmissions are required to be given due importance during the design of spectrum sensing framework and deriving the sensing schedule. The consideration of an acceptable level of QoS for SUs is important to enhance the overall network throughput in CR communication \cite{qossun2011optimal}. Similarly the issues of fairness in terms of different priority situations and users interest are to be considered during the design and development of schemes for sensing and scheduling. In presence of multiple channels, selection of SUs to perform spectrum sensing in fair manner is important for sensing accuracy and throughput optimization. Determination of fairness and adopting the technique into CSSS scheme depends on parameters like channel condition and distance of SUs from secondary base station and PU transmitters. If a SU is selected by CSSS policy which experiences bad channel condition on top of other SUs, it will result in poor throughput due to the fact that the SU will produce inaccurate sensing result. Similarly the SUs locating close to secondary base station will always experience better channel condition and will produce better spectrum sensing and data transmission results. Now if the scheduling algorithm is designed to be greedy to choose only such closely located SUs by the CSSS policy it may result in unfairness to all other SUs \cite{fairnesszhang2012proportional}. Because all the other SUs are unlikely to be given opportunity to take part in sensing and data transmission. Therefore, designing a sensing scheduling scheme for considering both QoS and fairness in MC-CRN is important.

\section{Methods for modelling CSSS schemes}
The approaches to design the CSSS problem can be dependent on the methods of modelling the problem which can be classified as follows:
\subsubsection{Game Theory}
Game Theory (GT) is a branch of mathematics which deals with the conflict and cooperation between players by considering the dynamic interaction amongst them. GT also takes into accounts the trade-off between the benefits and cost of strategies taken by players. For modeling the problem of CSSS, a game theory based approach has been proposed in the approach \cite{5li2017utility}. The objective of this approach is to improve the utility of the network along with a desired level of detection performance. The approach proposes two game theoretical model: first based on evolutionary game that decides which action is to be taken by SUs during CSS and second, based on coalitional game that decides which channels should be sensed by which SUs.  The first model consider that there may be some SUs who would like to participate in sensing operation are called as contributors and the others who do not want to participate in sensing but only overhear the sensing decisions by others are called as free-riders. This model considers the trade-off of the contributors revenue received and energy cost because of the participation in sensing and the free riders energy saving benefits and penalty cost because of not participation in sensing. The second model states that each contributor should join a coalition which brings the maximum information about the corresponding channel's status and thus increases the detection performance.
\subsubsection{Discrete Convex Analysis}
Discrete Convex Analysis is a paradigm for discrete optimization that combines the ideas in continuous optimization (the decision variables in the optimization problem are allowed to take continuous values) and combinatorial optimization (the decision variables in the optimization problem are allowed to take discrete values) to establish a unified theoretical framework for nonlinear discrete optimization. In \cite{6sun2013energy}, authors formulated the CSSS problem as a nonlinear binary programming problem with the objective to optimize throughput as well as energy consumption and adopt a three step approach to solve it. In the first step, the optimal number of SUs assigned to sense each channel is obtained using the framework of $M/M$-convex theory. In the second step, the SU assignment matrix is determined using the $L/L$-convex theory. This assignment matrix indicates which SU is assigned to which particular PU channel. In the third step, the optimal number of SUs participating in sensing for each channel is obtained based on the SU assignment scheme prepared in the second step. By combining these three steps, a complete and efficient SU assignment scheme for sensing in multiple PU channels is obtained.

\subsubsection{Outer Linearization}
The Outer Linearization (OL) method is a technique to solve the convex optimization problems which are non-linear in nature by capturing the global properties of the objective and constraints. In the work \cite{7eryigit2013energy}, the CSSS problem is formulated as a mixed integer non-linear programming problem (MINLP) and to solve this problem an OL based approximation algorithm is proposed which approximate the optimal solution up to a desired tolerance level (i.e. given by $\epsilon$ in this work). Using local information the OL method solves the optimization problem in an iterative fashion by relaxing and linearizing the constraint until a solution of required level is achieved. As discussed in \cite{7eryigit2013energy} , the OL algorithm starts with ignoring the mixed integer linear constraint of the formulated optimization problem to obtain an initial solution and then check the solution to test whether it satisfies all previously ignored constraints or not. If it satisfied those previously ignored constraints then the result is optimal otherwise the most violated constraint is linearized using the current solution and is added to the current problem as a new constraint to obtain another solution. This linearization process is repeated until all the constraints are satisfied with a $\epsilon$ tolerance.
\subsubsection{Dynamic Programming} 
Authors in \cite{9liu2016robust} formulated the CSSS problem as an optimization problem whose main objective is to maximize the total expected normalized throughput that might be achieved from all PU channels by deciding the optimal sensing strategy. A dynamic programming (DP) based solution is proposed in \cite{9liu2016robust} to decide a sensing strategy for SUs in order to find the optimal assignment such that throughput can be maximized. The stage of the DP is considered to be the channel numbers. At any channel, this algorithm decide how many of the remaining users should be assigned such that the optimal assignment can be obtained. The decision variables for this algorithm are:  the number of users, the instantaneous payoff of the  channel in which assignment is done, and the value function $v_k(n)$ which is the total expected throughput that can be obtained from the optimal assignment from now on when k channels and n users are still remain.

\subsubsection{Greedy Algorithm} 
Authors in \cite{8deng2012energy}, formulated the CSSS problem as an energy-efficient cooperative spectrum sensing problem (ECSSP) in which a series of non-disjoint subsets of SUs need to be decided from whole set of SUs with a given time co-efficient such that each subset satisfies the necessary detection and false alarm thresholds to maximize the network lifetime. The ECSSP problem is further modeled as a least weighted subset problem (LWSP) with the objective to find out a least weighted subset from the SUs set for a given weight co-efficient. Two greedy based algorithm have been proposed to solve LWSP problem: one General Greedy (GG) algorithm and the other as $\lambda$G algorithm. Both the algorithms work in similar manner by starting with an empty subset of SUs and then simply going through the SUs in decreasing order of efficiency and adding them one by one into the subset until it satisfies the required condition. The difference in these two algorithms lies on the way of deriving efficiency. In case of GG algorithm, efficiency is derived with the help of weight and constraint co-efficient while in case of $\lambda$G algorithm efficiency is also dependent on $\lambda$ along with weight and constraint co-efficient of the problem. In $\lambda$G algorithm, $\lambda$ is used to coordinate the proportions of weight and constraint coefficient during the sorting of the efficiency values.

\section{Future Directions}
Towards the development of efficient CSSS scheme for MC-CRN there are several open challenges which need to be addressed and investigated in future for success of CRN. Few of them are listed as below:

\subsection{Designing Distributed CSSS scheme for MC-CRN:}
Most of the existing works on CSSS in the literature assume a centralized base station to carry out all the scheduling tasks which incur extra overhead. But handling a reasonably large ah-hoc based CRN consisting of huge number of SUs over a larger coverage area is a complex problem. In such a situation designing a distributed CSSS scheme would be a better solution.
\subsection{Handling the Malicious users:}
Usually it is assumed that either SUs are rational or selfish but not malicious. But practically it might happen that some of the SUs may be compromised by some external unauthorized entities which may take part in sensing operation with bad intention. In presence of such an environment designing of CSSS scheme for MC-CRN would be an interesting problem.
\subsection{Joint prediction and sensing scheduling:}
Many a times PUs show temporal coherency within different licensed channels such that the appearance of PU on channels follow a pattern. In such situations a learning based prediction method may assist the decision making node to prepare an efficient scheduling strategy and can help to predict which channel to be sensed at what time to get better throughput. Developing a joint prediction and sensing scheduling scheme in this context is a challenging problem.
\subsection{Joint sensing and transmission scheduling:}
In MC-CRN the consideration of designing a transmission schedule by SUs alongside sensing schedule is beneficial for throughput optimization, because a SU should get benefit according to its contribution during sensing. In order to get better network performance along with the fairness among the SUs, designing of a joint sensing and transmission scheduling scheme play an important role.
\section{Conclusion}
In this article, we present a study on the problem of designing CSS scheduling scheme for multi-channel CRN highlighting the objectives, issues and challenges, methods and future directions. It is found that CSS scheduling for multi-channel CRN is important to be considered to improve detection performances as well as throughput and seamless communication. Various methods proposed in the literature to design and develop CSSS schemes are still primitive and need further investigation for their working in practical field. We summarized the discussion and identified several future research directions towards designing the efficient CSSS schemes for multi-channel CRN.


%
%

\bibliographystyle{IEEEtran}
\bibliography{IEEEabrv,ref}
\end{document}